\documentclass[cits]{PoS}

\title{Curvature of the phase transition line in the $\mu$-T plane}

\ShortTitle{Curvature of the phase transition line in the $\mu$-T plane}

\author{Zolt\'an Fodor,$^{abc}$ \speaker{Christa Guse},$^{a}$ S\'andor D. Katz$^{ab}$
        and K\'alm\'an K. Szab\'o$^{a}$\\
        \\
\llap{$^a$}Department of Physics, University of Wuppertal,\\
Gauss Strasse 20, D-42119, Wuppertal, Germany\\
\llap{$^b$}Institute for Theoretical Physics, E\"otv\"os University,\\
Pazmany 1, H-1117 Budapest, Hungary\\
\llap{$^c$}John von Neumann Institute for Computing (NIC),\\
DESY, D-15738, Zeuthen / FZJ, D-52425, Juelich, Germany\\

E-mail: \email{fodor@bodri.elte.hu}, \email{guse@physik.uni-wuppertal.de},
     \email{katz@bodri.elte.hu}, \email{szaboka@general.elte.hu}
}

\abstract{We determined the curvature of the phase transition line in the $\mu$-$T$ plane using a Taylor expansion in $\mu$. The Polyakov loop and the strange quark number susceptibility were measured to locate the pseudocritical line.
The analysis was carried out on $N_t=4,6,8,10$ lattices generated with a Symanzik improved gauge and stout-link improved (2+1) flavour staggered fermion action using physical quark masses.}

\FullConference{The XXV International Symposium on Lattice Field Theory\\
                 July 30 - August 4 2007\\
                 Regensburg, Germany}

\begin{document}

\section{Introduction}
The thermodynamics of strongly interacting matter at finite temperature has been studied extensively. It has been shown that the QCD transition is a cross-over \cite{Aoki:2006we}. 
This result is of importance for the physics of the early universe. It is also interesting to examine the QCD phase diagram at finite chemical potential $\mu_B \neq 0$ which would be relevant for the description of heavy ion collisions.
Lattice simulations in this regime have been hindered by the sign problem ever since the first attempts~\cite{Barbour:1986jf}. The sign problem originates from the fact that the fermion matrix $M$ loses its $\gamma_5$ hermiticity at finite chemical potential $M \neq \gamma_5 M^\dag \gamma_5$. Thus the fermion determinant $\det\left(M\right)$ becomes complex and spoils any method based on importance sampling. There are several different methods to circumvent this problem for small $\mu_B$ based on either performing simulations at imaginary $\mu_B$ and then analytically continuing the results to the physically interesting case of real $\mu_B$ \cite{Philipsen:2004bi} or reweighting samples produced with a real and positive $\det\left( M \right)$~\cite{Barbour:1997ej,Fodor:2001au}. For our calculations we used a reweighting technique, where all observables measured on the $\mu_B=0$ configurations are expanded in $\mu_B$.
The analysis was performed using the same configurations as in \cite{Aoki:2006br}.

\section{Reweighting and Taylor expansion}
The reweighting technique uses configurations generated at $\mu_B=0$ (where the fermion determinant is real and positive), to access observables $\mathcal{O}$ at finite chemical potential.
\begin{eqnarray}
\left< \mathcal{O} \right> &=& \frac{\int \mathcal{D} U \,\mathcal{O}\,\left[ \det\left( M(\mu_B) \right)^{n_f/4} \right]\,e^{-S_g(\beta)}}{\int \mathcal{D} U \,\left[ \det\left( M(\mu_B) \right)^{n_f/4} \right]\,e^{-S_g(\beta)}} \nonumber\\
&=&\frac{\int \mathcal{D} U\,\mathcal{O}\,\omega' \left[ \det\left(M(0)\right)^{n_f/4} \right]\,e^{-S_g(\beta_0)}}{\int \mathcal{D} U\,\omega' \left[ \det\left(M(0)\right)^{n_f/4} \right]\,e^{-S_g(\beta_0)}}=\frac{\left<\mathcal{O}\, \omega'\right>_0}{\left<\omega'\right>_0} \nonumber\\
&&\omega'= \frac{\det \left( M(\mu_B) \right)^{n_f/4}}{\det \left( M(0) \right)^{n_f/4}} \, e^{-(S_g(\beta)-S_g(\beta_0))}
\end{eqnarray}
The expectation value $\left< \mathcal{O} \right>$ at $\mu_B$ and $\beta$ is rewritten as the ratio of two expectation values at $\mu_B=0$ and $\beta_0$.
The chemical potential range where the reweighting is effective is limited by the decreasing overlap between the sample generated at $\mu_B=0$ and the target ensemble at $\mu_B \neq 0$.
As the distribution of the sample becomes narrower with increasing volume the overlap problem gets also more severe, thus requiring higher statistics when approaching the continuum limit.
It is quite expensive to calculate the determinant $\det \left(M(\mu_B)\right)$ exactly, especially for large lattices. For that reason we decided to use a Taylor expansion in $\mu$. This technique has been used several times, especially by the Bielefeld-Swansea Collaboration (for a recent paper with sixth order expansion see~\cite{Allton:2005gk}). For gluonic observables which do not depend on the chemical potential explicitly this expansion is very easy to perform.
\begin{eqnarray}
\left< \mathcal{O} \right> = \left< \mathcal{O} \right>_0 + 2 \,\left< \mathcal{O} O_1 \right>_0 \, \mu_q +\left[ \left< \mathcal{O} (2 O_1^2+O_2)\right>_0 - \left< \mathcal{O} \right>_0 \left< 2 O_1^2+O_2 \right>_0 \right] \mu_q^2
\end{eqnarray}
with
\begin{eqnarray}
O_1&=&\frac{1}{4} \left.\frac{\partial}{\partial \mu_q} \left[ \ln \det \left(M(\mu_q)\right) \right] \right|_{\mu_q=0} = \frac{1}{4} \mathrm{tr} \left( M^{-1} M' \right)
\end{eqnarray}
and
\begin{eqnarray}
O_2&=&\frac{1}{4} \left.\frac{\partial^2}{\partial \mu_q^2} \left[ \ln \det \left(M(\mu_q)\right) \right] \right|_{\mu_q=0} = \frac{1}{4} \mathrm{tr} \left( M^{-1} M'' - M^{-1} M' M^{-1} M' \right)
\end{eqnarray}
One can also note, that at $\mu_q=0$ the odd order derivatives of $\ln \det \left(M(\mu_q)\right)$ are purely imaginary and the even order derivatives are real. This follows from the fermion matrix identity:
\begin{eqnarray}
M^\dag \left(\mu_q\right) = \gamma_5 M\left(-\mu_q\right) \gamma_5
\end{eqnarray}
Therefore it is clear that for a real observable the linear term in the expansion is zero. 

\section{Observables}

All observables that we compute in this work namely the Polyakov loop and the strange quark number susceptibility do not depend explicitly on $\mu_B$ and can thus be expanded like gluonic observables.

\subsection{Polyakov loop}
The Polyakov loop is given by the trace over a product of gauge links along a line in time direction:
\begin{eqnarray}
P = \frac{1}{N_s^3} \sum_{\bf x} \mbox{tr}\left[ U_4\left( {\bf x}, 0 \right) U_4\left( {\bf x}, 1 \right) \dots  U_4\left( {\bf x}, N_t-1 \right) \right]
\end{eqnarray}
In pure gauge theory a change in the expectation value of the Polyakov loop to a non-vanishing value signals the spontaneous breakdown of the Z(3) symmetry and thus the onset of deconfinement. The Polyakov loop is also related to the quark-antiquark free energy at infinite separation:
\begin{eqnarray}
\left| \left< P \right> \right|^2 = \exp\left( - \Delta F_{q\bar{q}} \left( r \rightarrow \infty \right) \right)
\end{eqnarray}
Here $\Delta F_{q \bar{q}}$ denotes the difference of the free energies of the quark-gluon plasma with and without the quark-antiquark pair. It is then possible to renormalize the Polyakov loop by renormalizing the free energy of the quark-antiquark pair. For that purpose one uses the renormalization condition $V_R(r_0)=0$ to renormalize the potential at $T=0$ for each lattice spacing \cite{Aoki:2006br}. As the ultraviolet divergencies are the same at $T=0$ as at finite $T$ the same renormalization shift can then be used to renormalize the free energy. The renormalized Polyakov loop is given by
\begin{eqnarray}
\left| \left< P_R \right> \right| = \left| \left< P \right> \right| \, \exp\left( V(r_0)/(2T) \right),
\end{eqnarray}
where $V(r_0)$ is the unrenormalized $T=0$ potential obtained from Wilson-loops. The transition temperature can be defined as the peak in the temperature derivative of the Polyakov loop, that is the inflection point of the Polyakov loop curve.

We also note that the linear term in the Taylor expansion of the Polyakov loop is zero.
This can be shown by using the CPT symmetry \cite{Alles:2002wh}. It is enough to show that $\left< P \right>_\mu=\left< P \right>_\mu^*$. This follows from the CPT transformation of the expectation value of the Polyakov loop:
\begin{eqnarray}
\left< P \right>_\mu = \left< P \right>_\mu^{\tiny\mbox{CPT}} &=& \int \mathcal{D} U^{\tiny\mbox{CPT}} \,e^{-S_g(U^{\tiny\mbox{CPT}})} \det (M_\mu(U^{\tiny\mbox{CPT}})) \, P^{\,\tiny\mbox{CPT}} \nonumber\\
&=& \int \mathcal{D} U \,e^{-S_g(U)} \left[\det (M_\mu(U))\right]^* \,P^*=\left( \left< P \right>_\mu \right)^*
\end{eqnarray}

\subsection{Strange quark number susceptibility}
The strange quark number susceptibility $\chi_s$ is defined by
\begin{eqnarray}
\frac{\chi_s}{T^2} = \left. \frac{1}{TV} \frac{\partial^2 \log Z}{\partial \mu_s^2} \right|_{\mu_s=0}
\end{eqnarray}
It can be related to the event-by-event fluctuations in heavy-ion experiments. As $\chi_s$ has a well defined continuum limit, renormalization is not necessary. The transition temperature can be defined as the peak in the temperature derivative of the strange quark number susceptibility, that is the inflection point of the susceptibility curve.

As the strange quark number susceptibility is a real observable the linear term in the $\mu_B$ expansion vanishes as was shown above.

\section{Simulation Setup}
We used a Symanzik improved gauge and stout-link improved staggered fermionic lattice action in order to reduce taste violation~\cite{Aoki:2005vt}. The configurations were generated with an exact RHMC algorithm. We determined a line of constant physics (LCP) using physical masses for the light quarks $m_{u,d}$ as well as for the strange quark $m_s$. The LCP was fixed by setting $m_K/f_K$ and $m_K/m_\pi$ to their physical values \cite{Aoki:2006br}.
We used four different lattice spacings $N_t=4,6,8,10$ with an aspect ratio of $N_s/N_t=4$.
The scale was fixed by $f_K$ and its unambiguity checked by calculating $m_{K*}$, $f_\pi$ and $r_0$. The operators $O_1$ and $O_2$ were computed using a random noise estimator for the traces. The number of random vectors was chosen to give an error of the same size as the statistical error. It was between 50 and 160 in all cases.

\section{Results}

\begin{figure}[h]
\begin{minipage}{0.5\textwidth}
\include{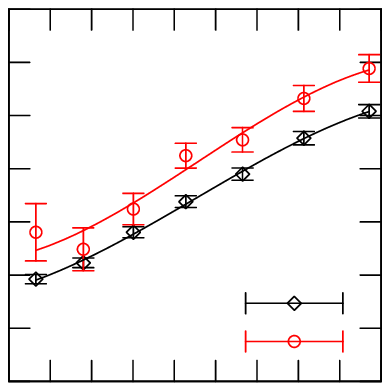}
\end{minipage}
\begin{minipage}{0.45\textwidth}
\include{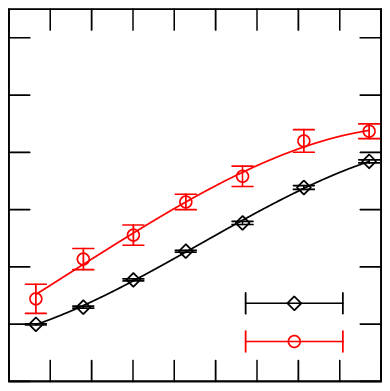}
\end{minipage}
\caption{On the left hand side the temperature dependence of the expectation value of the Polyakov loop is shown for $\mu\,a=0$ and $\mu\,a=0.1$. The graph on the right hand side depicts the temperature dependence of the strange quark number susceptibility for $\mu\,a=0$ and $\mu\,a=0.1$. The results were obtained on the $N_t=6$ lattices.}
\label{fig1a}
\end{figure}

Figure 1 shows the Polyakov loop and the strange quark number susceptibility for $\mu\,a=0$ and $\mu\,a=0.1$ at $N_t=6$.
We want to compute the curvature $\kappa$ of the phase transition line at $\mu_B=0$, which we define by
\begin{eqnarray}
T_c(\mu_B)=T_c \left( 1-\kappa \, \frac{\mu_B^2}{T_c^2} \right)
\end{eqnarray}
that is
\begin{eqnarray}
\kappa = - T_c \left.\frac{d T_c(\mu_B)}{d \mu_B^2} \right|_{\mu_B=0} \, .
\end{eqnarray}
Here it is assumed that the phase transition line can be well described by a purely quadratic term in $\mu_B$, whereas the fourth order term is negligible.
Now we use 
\begin{eqnarray}
\left.\frac{d T_c(\mu_B)}{d \mu_B^2} \right|_{\mu_B=0} = \left.\frac{\partial \mathcal{O}}{\partial\mu_B^2}\right|_{\mu_B=0}\left/\left.\frac{\partial \mathcal{O}}{\partial T}\right|_{T=T_c(0)}\right.
\end{eqnarray}
to directly extract the curvature from $\partial \mathcal{O}/\partial\mu_B^2$ and $\partial \mathcal{O}/\partial T$. Here $\mathcal{O}$ stands for the Polyakov loop or the strange quark number susceptibility respectively. Figure 2 shows $- T_c d T_c(\mu_B)/d \mu_B^2$ computed by $d T_c(\mu_B)/d \mu_B^2 = \partial \chi_s/\partial\mu_B^2\left( \partial \chi_s/\partial T \right)^{-1}$ for the $N_t=4$ lattices. We use an average over the points near $T_c(\mu_B=0)$ to extract the curvature at $T_c$ and can thus increase the statistics.

\begin{figure}[h]
\center
\include{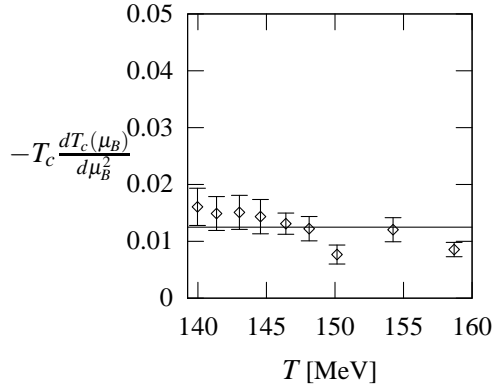}
\caption{This plot shows the $- T_c\,d T_c(\mu_B)/d \mu_B^2$ computed by $d T_c(\mu_B)/d \mu_B^2 = \partial \chi_s/\partial\mu_B^2\left(\partial \chi_s/\partial T\right)^{-1}$ for the $N_t=4$ lattices.}
\label{fig3}
\end{figure}

\begin{figure}[h]
\begin{minipage}{0.5\textwidth}
\include{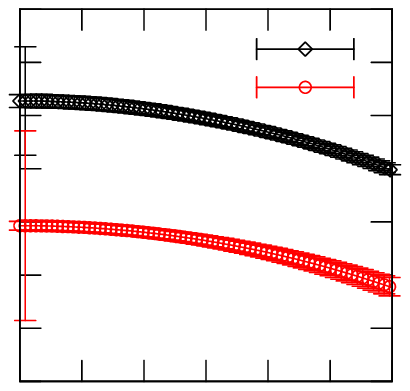}
\end{minipage}
\begin{minipage}{0.45\textwidth}
\include{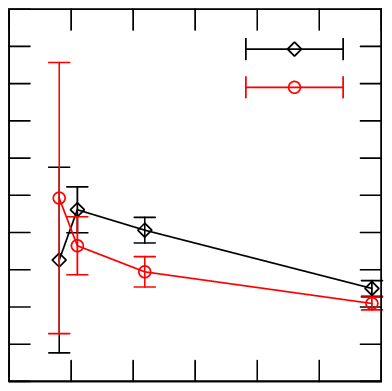}
\end{minipage}
\caption{The left plot shows the phasediagram in the $\mu_B$-$T$ plane for the lattice spacing $N_t=4$. The red curves give the crossover line computed by finding the inflection point of the Polyakov loop, the black one corresponds to the computations using the strange quark number susceptibility. The big errorbars at the start of the curve denote the total error of $T_c(\mu=0)$, the smaller errorbars include just the error of the shift $\Delta T$. As the phasediagram was extracted from a Taylor expansion it is only correct to leading order. On the right hand side the lattice spacing dependence of the curvature $\kappa$ is displayed. No safe continuum extrapolation is possible with our present statistics. Here the red and black lines correspond to the results obtained from the Polyakov loop and the strange quark number susceptibility respectively.}
\label{fig3a}
\end{figure}

In our case $\partial \mathcal{O}/\partial\mu_B^2$ is directly given by the quadratic term in the $\mu$ expansion. Whereas $\partial \mathcal{O}/\partial T$ can be extracted from a fit to the Polyakov loop and the quark number susceptibility near $T_c$.
For one point we also calculated $\partial \mathcal{O}/\partial\mu_B^2$ by exact determinant evaluation \cite{Fodor:2001pe} and found a perfect agreement.

We determine the error of the curvature by carrying out the whole analysis for different jackknife-samples and by varying the fit range. 

To extract the continuum limit of $\kappa$ one needs at least three points in the scaling region. It was established in \cite{Aoki:2006br} that our $N_t=4$ lattices aren't in this region. Thus we use the $N_t=6,8,10$ results for the extrapolation, see Figure 3. Unfortunately the present statistics for the $N_t=10$ lattices is very limited and thus the errors are large. It is an ongoing project to increase the statistics to give a reliable continuum value for the curvature.

\section{Summary}
We have computed the curvature $\kappa$ of the $\mu$-$T$ pseudocritical line at $\mu_B=0$. The simulations were performed at physical quark masses with $n_f=2+1$ flavours for 4 different lattice spacings. We used a Taylor expansion in $\mu_B$ of the Polyakov loop and the strange quark mass susceptibility and determined their inflection points in order to locate the crossover line. 

\acknowledgments{The computations were carried out on the 370 processor PC cluster of E\"otv\"os University, on the 1024 processor PC cluster of Wuppertal University, on the 107 node PC cluster equipped with Graphical Processing Units at Wuppertal University~\cite{Egri:2006zm} and on the Blue-Gene/L at FZ J\"ulich. We used a modified version of the publicly available MILC code \cite{MILC} with next-neighbour communication architecture for PC-clusters \cite{Fodor:2002zi}.
Partial support of grants of \hbox{DFG F0 502/1}, \hbox{EU I3HP} and \hbox{OTKA AT049652} is acknowledged.}


\begin{thebibliography}{99}
  \bibitem{Aoki:2006we} Y.~Aoki, G.~Endrodi, Z.~Fodor, S. D.~Katz, K. K.~Szabo, \emph{The order of the quantum chromodynamics transition predicted by the  standard model of particle physics}, \emph{Nature} {\bf 443} (2006) 675-678 [{\tt hep-lat/0611014}]
  \bibitem{Barbour:1986jf} I.~Barbour et. al., \emph{Problems with Finite Density Simulations of Lattice QCD}, \emph{Nucl. Phys.} {\bf B275} (1986) 296
  \bibitem{Philipsen:2004bi} Ph.~de Forcrand, O.~Philipsen, \emph{The QCD phase diagram for three degenerate flavors and small baryon  density} \emph{Nucl. Phys.} {\bf B673} (2003) 170-186 [{\tt hep-lat/0307020}]
  \bibitem{Barbour:1997ej} I. M.~Barbour, S. E.~Morrison, E. G.~Klepfish, J. B.~Kogut, M. P.~Lombardo, \emph{Results on finite density QCD} \emph{Nucl. Phys. Proc. Suppl.} {\bf 60A} (1998) 220-234 [{\tt hep-lat/9705042}]
  \bibitem{Fodor:2001au} Z.~Fodor, S. D.~Katz, \emph{A new method to study lattice QCD at finite temperature and chemical potential}, \emph{Phys. Lett.} {\bf B534} (2002) 87-92 [{\tt hep-lat/0104001}]
  \bibitem{Aoki:2006br} Y.~Aoki, Z.~Fodor, S. D.~Katz, K. K.~Szabo, \emph{The QCD transition temperature: Results with physical masses in the continuum limit}, \emph{Phys. Lett.} {\bf B643} (2006) 46-54 [{\tt hep-lat/0609068}]
  \bibitem{Allton:2005gk} C. R.~Allton et. al., \emph{Thermodynamics of two flavor QCD to sixth order in quark chemical potential}, \emph{Phys. Rev.} {\bf D71} (2005) 054508 [{\tt hep-lat/0501030}]
  \bibitem{Alles:2002wh} B.~Alles, E. M.~Moroni, \emph{A proposal for simulating QCD at finite chemical potential on the  lattice}, (2002) [{\tt hep-lat/0206028}]
  \bibitem{Aoki:2005vt} Y.~Aoki, Z.~Fodor, S. D.~Katz, K. K.~Szabo, \emph{The equation of state in lattice QCD: With physical quark masses towards the continuum limit}, \emph{JHEP} {\bf 01} (2006) 089 [{\tt hep-lat/0510084}]
  \bibitem{Fodor:2001pe} Z.~Fodor, S. D.~Katz, \emph{Lattice determination of the critical point of QCD at finite T and mu}, \emph{JHEP} {\bf 03} (2002) 014 [{\tt hep-lat/0106002}]
  \bibitem{Egri:2006zm} G. I.~Egri et. al., \emph{Lattice QCD as a video game}, \emph{Comput. Phys. Commun.} {\bf 177} (2007) 631-639 {[\tt hep-lat/0611022]}
  \bibitem{MILC} MILC Collaboration, \emph{public lattice gauge theory code}, \emph{see http://physics.indiana.edu/\~sg/milc.html}
  \bibitem{Fodor:2002zi} Z.~Fodor, S. D.~Katz, G.~Papp, \emph{Better than $1/$Mflops sustained: A scalable PC-based parallel computer for lattice QCD}, \emph{Comput. Phys. Commun.} {\bf 152} (2003) 121-134 [{\tt hep-lat/0202030}]
\end{thebibliography}
\end{document}